\documentclass[twocolumn,showpacs,preprintnumbers,amsmath,amssymb]{revtex4}
\usepackage{graphicx}
\usepackage{bm}
\usepackage{dcolumn}
\makeatletter
\newcommand{\rmnum}[1]{\romannumeral #1}
\newcommand{\Rmnum}[1]{\expandafter\@slowromancap\romannumeral #1@}
\makeatother

\begin{document}
\title{Unitarity potentials and  neutron matter at the unitary limit}
\author{Huan Dong, L.-W.\ Siu and T.\ T.\ S.\ Kuo}
\affiliation{Department of Physics and Astronomy,
Stony Brook University, New York 11794-3800, USA}
\author{R.\ Machleidt}
\affiliation{Department of Physics,
University of Idaho, Moscow, Idaho 83844, USA}
\date{}
\begin{abstract}
We study the equation of state  of  neutron matter
using a family of unitarity
potentials all of which are constructed to have infinite
$^1S_0$ scattering lengths $a_s$. For such system, a quantity of much
interest is the ratio
$\xi=E_0/E_0^{free}$ where $E_0$ is the true ground-state energy
of the system, and $E_0^{free}$ is that for the non-interacting system.
In the limit of $a_s\rightarrow
\pm \infty$, often referred to as the unitary limit,  this  ratio is expected
to approach a universal constant, namely $\xi\sim 0.44(1)$. 
In the present work we calculate
this ratio $\xi$ using a family of hard-core square-well  potentials 
whose $a_s$ can be exactly obtained, thus enabling us to have many 
potentials of different ranges and strengths, all with infinite $a_s$.
We have also calculated $\xi$ using a unitarity CDBonn potential obtained
by slightly scaling its meson parameters. The ratios $\xi$
given by these different unitarity potentials  are all close to each other
and also remarkably close to 0.44, suggesting that  
the above ratio $\xi$ is indifferent to the details of the
underlying interactions as long as they have infinite scattering length. 
A sum-rule and scaling constraint for the renormalized low-momentum interaction
in neutron matter at the unitary limit is studied.  The importance
of pairing in our all-order ring diagram and  model-space HF
calculations of neutron matter  is discussed.  
\end{abstract}
\pacs{pacs} \maketitle
\section{INTRODUCTION}

The `unitary limit' of an ultracold Fermi gas refers to the 
special scenario where the  inter-atomic interaction is
tuned to have its scattering length $a_s$ approaching
 infinity ($a_s>>k_F^{-1}$), leaving the Fermi momentum $k_F$ as the 
only relevant length scale in the many-body system. 
This limit was discussed as early as 1999 by Bertsch \cite{bertsch}, 
who challenged many-body 
theorists with the question ``What are the ground state properties of 
neutron matter, interacting with an infinite scattering length?". Under 
such condition the Fermions are strongly interacting, thus the determination 
of the ground state is highly non-trivial. However, the equation of state 
at this limit is expected
 to be an universal expression $E_0=\xi E_0^{free}$ where $\xi$ is an 
universal constant for any underlying Fermion systems. It is of much interest 
to determine $\xi$  experimentally or derive
it theoretically, many 
such attempts having been made
\cite{hara,gehm,partr,stewart,luo,
baker,heiselberg,carlson,carlson03,chang04,
perali,astrak,carlsonreddy,bulgac,schwenk05,nishida,lee06,hauss,chen,siu08,
borasoy,gezerlis,giorgini,epelbaum,abe,schaefer}. 
By far, the best numeric estimate on 
$\xi$ is considered to be  $\xi=$ 0.44(1) \cite{carlson},  
0.42(1) \cite{astrak} and 0.40(1) \cite{gezerlis}  according 
to quantum Monte Carlo methods. 
The scattering length of trapped ultracold Fermi gases 
 can in fact be magnetically tuned \cite{regal}
by way of the Feshbach resonance. 
This is indeed a very important achievement, and it is based on this that 
the above constant $\xi$ for trapped ultracold Fermi gases can be 
experimentally measured.
Several recently reported experimental values for $\xi$ are 
$0.46\pm0.05$ \cite{partr}, $0.46^{+0.12}_{-0.05}$ \cite{stewart}
and $0.39(2)$ to $0.435(15)$ \cite{luo}. 

In a previous work \cite{siu08}, we have calculated the constant $\xi$
for neutron matter using a tuned CDBonn realistic nucleon-nucleon (NN)
potential \cite{cdbonn}. Unlike the case of trapped Fermion gases, here 
 the scattering length of NN interactions still can not be tuned experimentally.
Thus we have chosen to tune the NN interaction by adjusting its meson
parameters, in line with the Brown-Rho scaling 
\cite{brownrho1} that the meson masses in nuclear medium
is suppressed compared with their in-vacuum values.
The $^1S_0$ scattering length of the original CDBonn potential is 
$a_s$=-18.9 {\rm fm}, which is already fairly long. We have found that to have 
a very large scattering length (such as -12070 {\rm fm})
the meson parameters of the CDBonn potential need to be adjusted 
only slightly (about $2\%$) \cite{siu08}. This tuned CDBonn potential actually
has given $\xi$ quite close to 0.44 for a wide range of densities
($\sim 0.02$ to $\sim 0.09$ {\rm fm}$^{-3}$) \cite{siu08}.

In the present work, we would like to investigate mainly the 
following question: Is the constant $\xi$ given by  other unitarity
potentials, defined as those with infinite scattering lengths,
also close to 0.44? This constant is supposed to be an universal
constant, and then its value should be independent of the detail
structure of the potentials as long as they have infinite scattering length.
In other words, $\xi$ given by all other unitarity potentials should
in principle be the same. It may be difficult to prove this analytically.
Before one can do so, it should be useful and of interest 
to first check this universality property
numerically as we shall do in the present work. As it is rather involved to
tune the various realistic NN potentials \cite{nijmegen,argonne,chiralvnn} 
to infinite scattering length,
 we have chosen to calculate $\xi$ from a family of
simple hard-core square-well (HCSW) potentials 
\begin{eqnarray}
 V(r)&=& V_c;~~ r < r_c \nonumber \\
     &&  V_b;~~ r_c<r < r_b \nonumber \\
     &&  0; ~~ r>r_b.
\end{eqnarray}
An advantage of this type of potentials
is that their scattering lengths can be given analytically and
thus exact unitarity potentials can be readily obtained. 
The above HCSW potentials are clearly very different from the
realistic NN potentials, yet as we shall discuss later the constant
given by various HCSW potentials are all amazingly close to 0.44,
as is also given by the tuned CDBonn potential.

In the following we shall present some details of our calculations.
In section IIA, we shall first briefly outline the renormalized low-momentum 
interaction $V_{low-k}$ 
\cite{bogner01,bogner02,coraggio,schwenk02,bogner03,holt}
on which our calculations will be based. We shall carry out calculations
using both HCSW and CDBonn potentials, both having strong repulsive cores.
As is well-known, these hard-core potentials are not suitable for being
used directly in many-body calculations. In contrast, the low-momentum
interaction $V_{low-k}$ is a smooth
and energy-independent potential, and it  has been extensively applied to
calculations of nuclear matter \cite{siu09}, nuclear structure \cite{corag08} 
and neutron star \cite{dong09}. This interaction is obtained   
using a renormalization group (RG) method where the high-momentum
components of an underlying potential beyond a decimation scale 
$\Lambda$ are integrated out. The low-energy phase shifts are preserved
by the RG procedure and so is the scattering length. In section IIB
the procedures for constructing our model unitarity potentials are
presented. Analytical expressions for the scattering length of HCSW
potentials are given. From these expressions, one can construct an unlimited
number of unitarity potentials (which all have infinite scattering length).

In section IIC, we shall describe how we calculate the neutron matter
equation of state (EOS) from the low-momentum interaction $V_{low-k}$. 
Two methods will be employed: a ring-diagram method \cite{siu08,siu09}  
and a model-space Hartree-Fock (MSHF) method.
 There have been  several many-body methods for calculating the  
EOS for large (number of particles approaching infinity) 
Fermionic systems such as atomic gases and nuclear matter.
 For a long time,
the Brueckner-Hartree-Fock (BHF) theory \cite{bethe,mach89,jwholt}
was the commonly used framework for such EOS calculations; 
an inconvenience of this method is that the BHF \textit{G}-matrix 
interaction is 
energy dependent, adding numerical complications. We shall use the
energy independent $V_{low-k}$ interaction mentioned earlier for
the EOS calculations, to circumvent the above inconvenience.
With the use of the $V_{low-k}$ interaction, the MSHF method provides a simple
way for calculating the constant $\xi$. In fact MSHF is rather similar
to BHF: In BHF the particle-particle correlations above $k_F$ are
incorporated by way of the energy-dependent \textit{G}-matrix, while in MBHF
they are included via the energy-independent $V_{low-k}$ interaction.
As we shall discuss later, the constant $\xi$ given by the ring-diagram
method and the MSHF method are in fact nearly identical 
(both close to 0.44). We shall also discuss that the MSHF mean field
potential should obey certain special constraint at the unitary limit.
 The importance of pairing in fermionic systems such as neutron matter 
 has been well known and extensively studied 
(see e.g. \cite{giorgini,schaefer} and 
references quoted therin). In section IIC, we shall  discuss that 
pairing has also played an important role in our all-order ring diagram 
calculations where a pair of  fermions are allowed
to interact any number of times, forming a quasi-boson composed
of a coherent pair of fermions. 
Our results will be presented and discussed in section III. A summary 
and conclusion will be given in section IV.

\section{FORMALISM}
\subsection{Low-momentum interactions} 
Since our  calculations are largely dependent on the recently 
developed low-momentum interaction $V_{low-k}$, it may be useful to
present first a short review on this interaction. 
As we know, the strong short-range repulsion contained in 
$V_{NN}$ usually requires a special renormalization treatment 
before being used  in many-body
 calculations. A familiar such treatment is the BHF \textit{G}-matrix method
where the \textit{G}-matrix interaction is obtained 
by summing the particle-particle 
ladder diagrams to all orders \cite{bethe,mach89,jwholt}. An inconvenient
aspect of this method is  the energy dependence of \textit{G}, 
making the many-body
calculations based on \textit{G} rather complicated.
In the past several years, there has been much progress in the 
 RG approach to 
the NN interaction. A central idea here is that to describe the low-energy
properties of a physical system it should be adequate to employ only 
a low-momentum effective interaction confined within a momentum decimation 
scale $\Lambda$.  $V_{low-k}$ is such a low-momentum effective interaction.
Starting from a realistic NN potential $V_{NN}$, $V_{low-k}$
is obtained from the following \textit{T}-matrix equivalence equations
\cite{bogner01,bogner02,coraggio,schwenk02,bogner03,holt}:
The half-on-shell \textit{T}-matrix for the full-space interaction is
\begin{multline}
T(k',k,k^2)=V_{NN}(k',k) \\
+\frac{2}{\pi}\mathcal{P}\int_0^\infty
\frac{V_{NN}(k',q)T(q,k,k^2)}{k^2-q^2}q^2dq,
\end{multline}
where $\mathcal{P}$ denotes principal-value integration 
and the intermediate state momentum $q$ is integrated 
from $0$ to $\infty$ covering the whole space. 
Then we define a low-momentum half-on-shell \textit{T}-matrix by
\begin{multline}
T_{low-k}(k',k,k^2)=V_{low-k}(k',k)\\
+\frac{2}{\pi}\mathcal{P}\int_0^\Lambda
\frac{V_{low-k}(k',q)T_{low-k}(q,k,k^2)}{k^2-q^2}q^2dq,
\end{multline}
where the intermediate state momentum is integrated from $0$ to $\Lambda$, 
the model space cut-off. The low momentum $V_{low-k}$ is then obtained by
requiring the equivalence of \textit{T}-matrix in the model space,
\begin{equation}
 T(k',k,k^2)=T_{low-k}(k',k,k^2); (k',k) \leq \Lambda.
\end{equation}
The low-momentum interaction $V_{low-k}$ is obtained by solving the above
three equations. Since this procedure preserves 
the half-on-shell \textit{T}-matrix, it certainly 
preserves the low-energy ($<\Lambda ^2$) phase shifts and
 scattering length $a_s$ of $V_{NN}$. This ensures that if $V_{NN}$ 
is tuned to be a unitarity
potential ($a_s \rightarrow \pm \infty$), so is the 
corresponding  $V_{low-k}$.

\subsection{Hard-core square-well unitarity potentials}
There are several realistic  NN potentials (CDBonn\cite{cdbonn},
Argonne\cite{argonne}, Nijmegen\cite{nijmegen}, Chiral\cite{chiralvnn})
which all describe the two-nucleon low-energy experimental data
very accurately. It would be of interest to obtain unitarity potentials
 from them by tunning their interaction 
parameters  slightly. But technically this is not easy to carry out.
 So far only the CDBonn potential has been tuned to
attain this limit ($a_s$=-12070 {\rm fm})\cite{siu08}. 
In the present work, we choose hard-core square-well (HCSW) potentials 
as given by Eq.(1) for 
further studying the properties of neutron matter at unitary limit. 
An advantage of using them is that their scattering length $a_s$ can be 
analytically obtained, allowing us to study 
the properties of neutron matter using  many HCSW potentials with any chosen
 scattering lengths (including infinity). 

 As indicated in Eq.(1), the HCSW potential is characterized by the
height $V_c$ of the repulsive core, the depth $V_b$ of its attractive well,
and the respective ranges $r_c$ and $r_b$. In the present work, we consider
neutron matter with interactions only in the  $^1S_0$ channel, whose 
phase shift $\delta$ is readily obtained from Eq.(5) and (6). Namely
\begin{eqnarray}
tan(\delta+K_3r_b)=\frac{K_3}{K_2}tan(K_2r_b+\alpha) \\
tan(\alpha+K_2r_c)=\frac{K_2}{K_1}tanh(K_1r_c)
\end{eqnarray}
with
\begin{eqnarray*}
K_1=\sqrt{(V_c-E)\frac{m}{\hbar^2}},\\
~~K_2=\sqrt{(E-V_b)\frac{m}{\hbar^2}},\\
~~K_3=\sqrt{E\frac{m}{\hbar^2}},
\end{eqnarray*}
where $E$ is the scattering energy in the center-of-mass frame. 
From the above results, the scattering length $a_s$
is obtained from a low energy expansion of $kcot\delta$ as
\begin{equation}
 a_s = -\frac{B}{A}
\end{equation} 
with
\begin{eqnarray} 
 A &=& K_{10}K_{20}-K_{20}^2tanh(K_{10}r_c)tan[K_{20}(r_b-r_c)], \nonumber \\
 B &=& K_{20}tanh(K_{10}r_c)+K_{10}tan[K_{20}(r_b-r_c)]-r_bK_{10}K_{20} 
\nonumber \\
   && +r_bK_{20}^2tanh(K_{10}r_c)tan[K_{20}(r_b-r_c)],         
\end{eqnarray}
where
\begin{equation*}
K_{10}=\sqrt{V_c\frac{m}{\hbar^2}},
~~K_{20}=\sqrt{-V_b\frac{m}{\hbar^2}}.
\end{equation*}
The effective range $r_e$ for this potential can also be derived analytically;
the result is not presented here because it is fairly lengthy.

  Eq.(7) implies that the condition for being a unitarity potential
(infinite scattering length) is \textit{A}=0, namely
\begin{equation}
r_b-r_c=\frac{1}{K_{20}}tan^{-1}[\frac{K_{10}}{K_{20}tanh(K_{10}r_c)}].
\label{width}
\end{equation}
It is well known that when the potential is tunned to the unitary limit
(the Feshbach resonance), it has a bound state with its energy 
approaching to zero. It is readily checked that  the
 condition for having such a bound state is the same as the  \textit{A}=0
one given above.
\begin{table}
\caption{Three different unitarity HCSW potentials.}
\centering
\begin{tabular}{c c c c c c c}
\hline\hline
Potentials & $V_c$[{\rm MeV}] & $r_c$[{\rm fm}] & $V_b$[{\rm MeV}] & $r_b$[{\rm fm}] & $a_s$[$\times10^6${\rm fm}] & $r_{e}$ [{\rm fm}]\\
\hline
HCSW01     & 3000 & 0.15 & -20 & 2.31 &15.2 & 2.36\\ 
HCSW02     & 3000 & 0.30 & -30 & 2.03 &3.38 & 2.21\\ 
HCSW03     & 3000 & 0.50 & -50 & 1.81 &-4.58 & 2.20\\ 
\hline
\end{tabular}
\label{table:hcsquare}
\end{table}

 The above condition  enables us to construct any number of
unitarity potentials by varying any three of the four parameters
$V_c$, $V_b$, $r_c$ and $r_b$. Some sample such potentials are listed
in Table  \Rmnum{1}.
As seen, they all have very large (infinite) scattering lengths 
 while  the potentials themselves are significantly 
different from each other. For example,
the depth of the attractive parts of them changes from -20 to -50 {\rm MeV}.
 The effective ranges of these potentials are also listed, all being
close to the unitarity-CDBonn's $r_e$ of $2.54 fm^{-1}$. 
These potentials will be used to calculate the universal
ratio $\xi\equiv E_0/E_0^{free}$ for neutron matter. As we 
shall soon discuss,
the ratios $\xi$ given by these three largely different potentials
are in fact nearly identical to each other.

\subsection{Ring-diagram and  model-space Hartree-Fock methods}

\begin{figure}[here]    
\scalebox{0.4}{\includegraphics{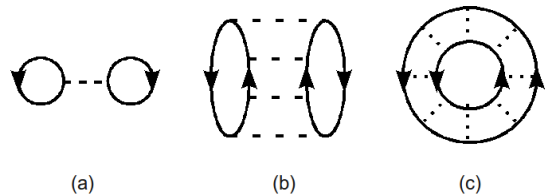}}
\caption{Diagrams included in the all-order \textit{pphh} ring-diagram 
summation for the ground state energy shift of nuclear matter.
Each dashed line represents a $V_{low-k}$ vertex.}
\end{figure}

We use the  ring-diagram method \cite{siu08,siu09} to calculate the
neutron matter EOS. In this method, the ground-state energy shift $\Delta E$ 
is given by the all-order sum of the \textit{pphh} ring diagrams as illustrated
in Fig.1, where (a), (b) and (c) are respectively 1st-, 4th- and 8th-order
such diagrams. ($\Delta E_0$ is defined as ($E_0-E_0^{free}$) where $E_0$
is the  true ground state energy and $E_0^{free}$ that for the non-interacting
system.) In this way, we have 
\begin{multline}\label{eng}
\Delta E_0=\int_0^1 d\lambda
\sum_m \sum_{ijkl<\Lambda}Y_m(ij,\lambda) \\ \times Y_m^*(kl,\lambda) \langle
ij|V_{low-k}|kl \rangle,
\end{multline}
where the transition amplitudes are 
\begin{equation}
Y_m^*(kl,\lambda)=\langle \Psi_m(\lambda,A-2)|a_l
a_k|\Psi_0(\lambda,A)\rangle,
\end{equation}
where $\Psi_0(\lambda,A)$ denotes the true ground state of nuclear matter
which has \textit{A} particles while
 $\Psi_m(\lambda,A-2)$ the $m$th true eigenstate of the (\textit{A}-2) system.
Note that $\lambda$ is a strength parameter, integrated from
0 to 1. 
Using the above $\Delta E_0$, the ratio $\xi$ is readily
obtained, namely $\xi$= 1 +$\Delta E_0/E_0^{free}$.

 By summing the $pphh$ ring diagrams to all orders, the above
amplitudes $Y$
 are given by the coupled equation of the form
\begin{eqnarray}
 AX+BY&=&\omega X \nonumber \\
 A^*Y+B^*X&=&-\omega Y,
\end{eqnarray}
where $X$ is $Y$'s RPA partner.\cite{song87}
 $\omega$ is the excitation energy corresponding to either $(E_m(A-2)-E_0(A))$
or $(E_m(A+2)-E_0(A))$. It is readily recognized  that 
the above RPA equation is of the same form as the well-known 
quasi-boson  RPA equations
\cite{rowe71}, 
resulting from  treating the many-body system approximately 
as a collection of quasi bosons (each composed of a pair
of interacting fermions). Thus our all-order ring-diagram calculation 
is actually a quasi-boson approximation for  neutron matter, 
treating it as a system of quasi bosons.
 (This is not unexpected
as RPA is a quasi-boson approximation.)
Our all-order ring-diagram calculation is  practically  a pairing 
calculation in the sense that we include only 
the boson-forming ring diagrams to all orders (both forward and 
backward ladder diagrams for a pair of fermions interacting 
any number of times), allowing them to form a coherent pair (quasi-boson).
 Note, however, that our calculation
becomes HF if we include only the 1st-order ring diagram,
because in this case the above
$Y$ amplitude becomes $Y(ij)=n_in_j$ where $n_i$=1 for $i<k_F$ and =0
otherwise. 
 Since we include only the
$^1S_0$ interaction,  the quasi bosons in our present ring-diagram calculation
are in fact all BCS-pairing bosons ($^1S_0$).

 The  ring-diagram method described above is a model-space formalism, 
where all nucleons are confined
within a momentum model space $P(k<\Lambda)$, and the decimation scale for
the $V_{low-k}$ interaction is the same $\Lambda$. The expression Eq.(10)
for the energy shift
$\Delta E_0$ is  rather complicated, and  is not
convenient for  studying the properties of  the underlying interaction
 at the unitary limit. To have a clearer way to study these properties, 
we have considered a simpler method to calculate $\Delta E_0$, namely a 
model-space HF method (MSHF). In this method, we have
\begin{equation}
\Delta E_0=\frac{1}{2}\Sigma _{k_1,k_2\leq k_F}
           \langle \vec k_1 \vec k_2|V_{low-k}^{k_F}|\vec k_1 \vec k_2 \rangle.
\end{equation}
This is also a model-space approach where all nucleons are
confined within a model space $P(k<k_F)$, but
MSHF has a special feature that the 
interaction used is renormalized according to the same $P(k<k_F)$,
namely the $V_{low-k}$ interaction is calculated
 with $\Lambda = k_F$ as denoted by $V_{low-k}^{k_F}$. A main difference
between MSHF and the usual HF calculations is that the interaction used
in the latter is not required to be renormalized according to the above model
space while it is so for the former.
 In fact MBHF is equivalent to the ring-diagram
method when one chooses $\Lambda$ to be its smallest value
allowed by the Pauli principle, namely $k_F$. 
When $\Lambda=k_F$, diagrams like (b) and (c) of Fig.1 no longer
exist (since particles with $k>k_F$ are disallowed in the model space);
 only diagram (a) remains, which is just MBHF. It may be mentioned that
MBHF is rather similar to the familiar BHF method \cite{bethe,mach89,jwholt}.
 They  both employ the same model space  $P(k<k_F)$ (all nucleons 
confined within the Fermi sea),
but they employ different model-space effective interactions.
In BHF the energy-dependent \textit{G}-matrix interaction is used while in MBHF 
  the energy-independent $V_{low-k}$ interaction is employed; both include
renormalizations from all-order particle-particle correlations beyond
$k_F$, but for \textit{G} the renormalization is carried out 
using an energy-dependent
method  while an energy-independent one is applied for $V_{low-k}$.   
The MBHF ground-state energy is given  by the sum of the kinetic
energy and  a simple integral, namely
\begin{eqnarray}
\frac{E_0}{A}&=&\frac{3}{5}\varepsilon_F+\frac{8}{\pi }
\int_0^{k_F}k^2dk  [1-\frac{3k}{2k_F}+\frac{k^3}{2k_F^3}]
\nonumber \\
&& \times \sum_{\substack{\alpha}} (2J_{\alpha}+1)
\langle  \alpha,k|V_{low-k}^{k_F}| \alpha,k \rangle.   
\label{binding}
\end{eqnarray}
In the above $\varepsilon _F$=$\frac{\hbar^2k_F^2}{2m}$,
 $k$ denotes the two-nucleon relative momentum and $J_{\alpha}$
is the total angular momentum for the partial wave $\alpha$. (We consider 
neutrons with interactions only in the $^1S_0$  partial wave, and
thus the above summation has only one term.) 
 The above result may
be useful for studying the properties of the $V_{low-k}$ interaction at the
 unitary limit. If the ratio
$E_0/E_0^{free}$ is equal to a constant $\xi$ at this limit, then the 
above equation implies that
$V_{low-k}^{k_F}$ at this limit must satisfy 
\begin{eqnarray}
\frac{3\varepsilon_F}{5}(\xi-1)&=&\frac{8}{\pi }
\int_0^{k_F}k^2dk  [1-\frac{3k}{2k_F}+\frac{k^3}{2k_F^3}]
\nonumber \\
&& \times
\langle ^1S_0,k|V_{low-k}^{k_F}|^1S_0,k\rangle.   
\label{constraint}
\end{eqnarray}
This is a rather strong constraint for the interaction.
Clearly there are many potentials which can satisfy
the above constraint, allowing many potentials to have the same ratio $\xi$.
In the following section, we shall discuss and numerically check 
this constraint.

Within the above MSHF framework, the single-particle (s.p.) potential $U$
is given  \cite{song87} as
\begin{eqnarray}
U(k_1)&=&\sum_{\substack{\alpha}}(2J_{\alpha}+1) \{ \frac{16}{\pi }
\int_0^{k_-}k^2dk \langle k \alpha|V_{low-k}^{k_F}|k \alpha \rangle  \nonumber \\
&& +\frac{2}{k_1 \pi }
\int_{k_-}^{k_+}kdk[k_F^2-k_1^2+4k(k_1-k)] \nonumber \\
&& \times \langle k \alpha|V_{low-k}^{k_F}|k \alpha \rangle \}
\label{potential}
\end{eqnarray}
with
\begin{equation*}
 k_-=(k_F-k_1)/2, ~~
 k_+=(k_F+k_1)/2.
\end{equation*}
The MSHF s.p. spectrum is 
\begin{equation}
\varepsilon(k_1)=\frac{\hbar^2k_1^2}{2m}+U(k_1)
\end{equation}
which can be well  approximated by 
\begin{equation}
\varepsilon(k_1)=\frac{\hbar^2k_1^2}{2m^*}+\Delta
\end{equation}
where $m^*$ is the effective neutron mass in medium and $\Delta$ is 
a constant related to 
the depth of the potential well. The MSHF ground-state energy can be expressed
in terms of $m^*$ and $\Delta$. Then if this energy is equal to
$\xi E_0^{free}$ at the unitary limit, the MSHF $m^*$ and $\Delta$ should
satisfy the linear constraint 
\begin{equation}
\xi=\frac{1}{2}+\frac{m}{2m^*}+\frac{5\Delta}{6\varepsilon_F}.
\label{ratioformula}
\end{equation}
In the following section, we shall also discuss and numerically check 
the above linear constraint concerning  the MSHF sp potential.

\begin{figure}[h]
\scalebox{0.42}{
\includegraphics[angle=-90]{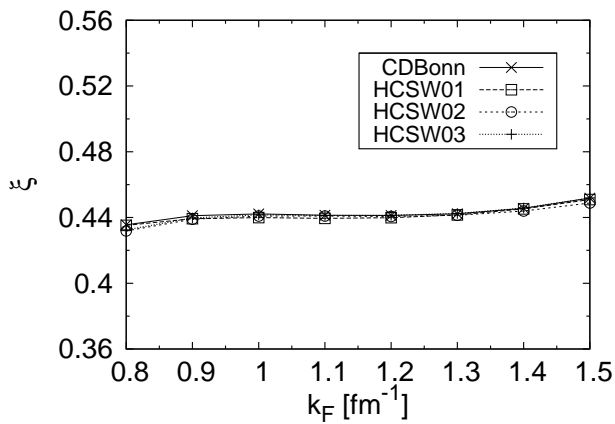}
}
\caption{\label{ringratio}Ring-diagram unitary ratios $\xi$ given 
by different unitarity potentials.}
\end{figure}    

\section{ RESULTS and DISCUSSIONS }
We first calculate the ratio $\xi$ with the ring-diagram method (see Eq.(9))
using both the three HCSW unitarity potentials of Table \Rmnum{1} and 
the unitarity CDBonn potential ($a_s$=-12070 {\rm fm}) \cite{siu08}. 
As shown in Fig.2, the results given by these
four potentials are all quite close to each other and to 0.44. 
A common decimation scale of $\Lambda=2.1$ {\rm fm}$^{-1}$ has been used
for the results presented there.  
We have found  that the $\xi$ values given by the ring-diagram calculations 
using the HCSW potentials are rather stable (variation less than $\sim 0.002$) 
with  $\Lambda$ in the range between $\sim2.0$ and $\sim 2.4$ {\rm fm}$^{-1}$.
As found in \cite{siu08}, they are also quite stable for the unitarity
CDBonn potential in the same range. Since these potentials have given
nearly identical results for the ration $\xi$, it is of interest 
to compare their $V_{NN}(k,k)$ matrix elements from which the $\xi$s
are calculated. Such a comparison is presented in Fig.3, and as seen 
they are actually quite
different. It is indeed surprising that the ratios $\xi$ given by these
vastly different interactions are nearly the same.  To illustrate the
key role of the unitary limit, we further calculate the ratios $\xi$ near 
the unitary limit as displayed in Fig.4. 
When $a_s$ is away from the unitary limit, 
the $\xi$ values given by these potentials are noticeably different, 
but they converge to a common value only as $1/a_s$ approaching zero.
Above results strongly suggest that the ratio $\xi$ is indifferent to 
the details of the underlying potentials, 
as long as they have infinite (or very large)
scattering lengths, namely they are all unitarity potentials.

\begin{figure}[tbh]
\scalebox{0.42}{
\includegraphics[angle=-90]{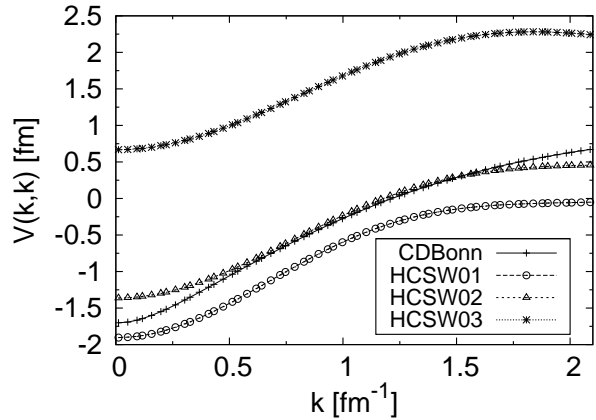} 
}
\caption{\label{vbarelemt}Diagonal matrix elements $V_{NN}(k,k)$ of different unitarity potentials.}
\end{figure}

\begin{figure}[h]
\scalebox{0.42}{
\includegraphics[angle=-90]{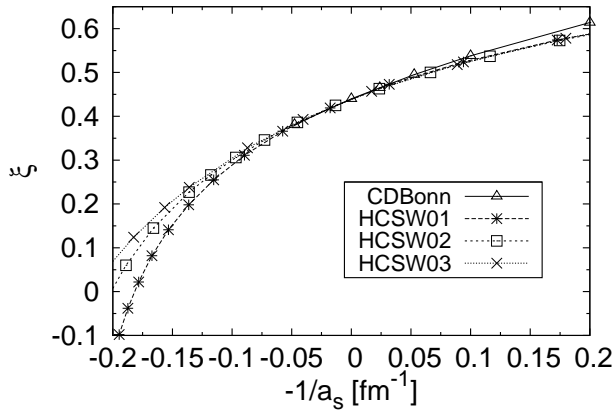}
}
\caption{\label{diffas} The ring-diagram unitary ratios $\xi$
near the unitary limit. 
$k_F=1.2$ {\rm fm}$^{-1}$ is used.}
\end{figure}
  Using the above four unitarity potentials, we have also calculated
the ratios $\xi$ using the MSHF method (see Eq.(14)). The results are
presented in Fig.5, and as seen the results given by the four potentials
are nearly identical and they are all quite close to 0.44
over a wide range of densities. It is of interest to note that
the MSHF results shown are remarkably close to the ring-diagram ones
of Fig.2. This close agreement is an indication that 
the hole-hole correlation diagrams like diagram (c) of Fig.1  may not
be  important for pure neutron matter. The contribution from the 
particle-particle correlations
 like diagrams (a) and (b) is included in both ring diagram and
MSHF calculations, but diagrams like (c) are included only in the former.
It may be noted that in both Fig.2 and Fig.5 our results are nearly 
constant
for  $k_F$ between $\sim 0.9$ and $\sim 1.4$ {\rm fm}$^{-1}$ (corresponding to
density range $\sim 0.016$ to $\sim 0.092$ {\rm fm}$^{-3}$). Outside this range,
small variations of $\xi$ start to appear. 
These variations are an indication that the methods we have employed 
for the calculation is not adequate for the high and very low 
density regions indicated above. So far we have employed only two-body 
$V_{low-k}$ interactions in the calculation. Three body interactions 
have not been considered in the present work, and their effects
 may be important. Further studies in this direction are needed.

\begin{figure}[h]
\scalebox{0.42}{
\includegraphics[angle=-90]{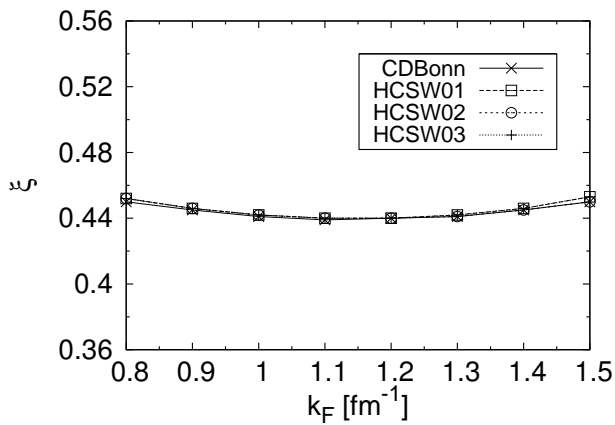}
}
\caption{\label{vlowkratio} MSHF unitary ratios $\xi$ given 
by different unitarity  potentials.}
\end{figure}

\begin{figure}[ht]
\scalebox{0.42}{
\includegraphics[angle=-90]{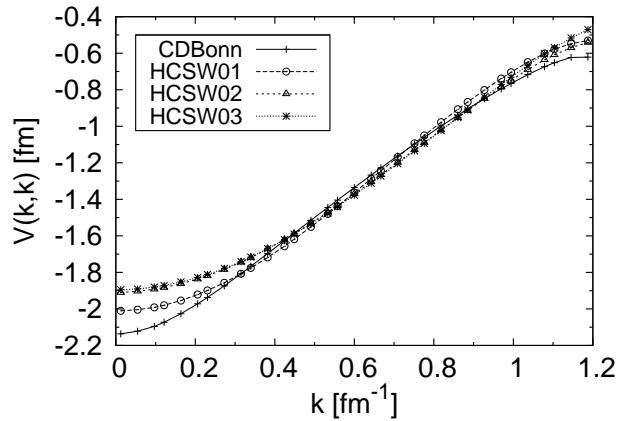}
}
\caption{\label{vlowkelemt}Diagonal matrix elements of 
$V_{low-k}^{k_F}$ from different unitarity potentials. 
$\Lambda=k_F=1.2$ {\rm fm}$^{-1}$ is used.}
\end{figure}

 In the MSHF method, the ratio $\xi$ is calculated from  the model-space 
effective interaction
$V_{low-k}^{k_F}$. We have seen from Fig.~5 that the ratios $\xi$ given by the
four different potentials are nearly the same. Does this mean that their
$V_{low-k}^{k_F}$ are also nearly the same?  In Fig.6 the matrix elements
of this interaction for four different unitarity potentials are presented.
As seen, they are clearly not identical to each other; there are 
significant differences among them. Eq.(14) is in fact
a strong constraint for the above interactions. To have $\xi$ being a
universal constant, the left-hand side of Eq.(14) is proportional to $k_F^2$.
Consequently the integral in the right-hand side must be equal 
to $C\times k_F^2$, $C$ being a constant,
 noting that its integrand and integration limit are both dependent
on $k_F$. This implies that at the unitary limit the $V_{low-k}^{k_F}$ 
interaction must satisfy certain stringent
requirements so as to satisfy the above constraint.

Similar to the effective interactions commonly used in effective
field theories \cite{epelb99,holt04}, we have found that our 
$V_{low-k}^{k_F}$ interactions can be highly accurately 
 simulated by 
low-order momentum expansions of the form 
\begin{equation}
\langle k| V_{low-k}^{k_F}|k \rangle = V_0 +V_2 (\frac{k}{k_F})^2 
+V_4 (\frac{k}{k_F})^4,
\end{equation}
where $V_0$, $V_2$ and $V_4$ are constants (independent of $k$ but dependent
on $k_F$). (The rms deviations for fitting the results of Fig.6 by
the above expansion are $\sim 2\times 10^{-2}$ and $\sim 2\times10^{-3}$ 
respectively for the CDBonn and HCSW cases, the fitting being very good.)
In terms of these constants, Eq.(14) assume a rather simple form, namely 
\begin{equation}
\frac{3\pi }{10}(\xi-1)=
k_F(\frac{V_0}{3}+\frac{V_2}{10}+\frac{3V_4}{70}).
\end{equation}
The above is an interesting sum-rule and scaling constraint, namely 
at the unitary
limit the strength sum ($V_0/3+V_2/2+3V_4/70$) is a constant for any
$k_F$ (density) and, in addition, it scales with $1/k_F$ (i.e. proportional
to $1/k_F$).
 In Table \Rmnum{2} we present some sample results, to check how well
do they satisfy the above constraint. As seen our results satisfy this
constraint very well. For each $k_F$, the values of the above sum
given by the four different unitarity potentials are all quite close
to each other. The values of this sum for different $k_F$ are different
but they are  all giving $\xi$ close to 0.44, in close agreement with
the $1/k_F$ scaling.

\begin{table}[h]
\caption{Low-order momentum expansion of $V_{low-k}^{k_F}$. Listed are the
coefficients $V_0$, $V_2$ and $V_4$ of Eq.(20), with the sum
$(V_0/3+V_2/10+3V_4/70)$ denoted as Sum. Four unitarity potentials
are used.}
\centering
\begin{tabular}{c c c c c c c}
\hline\hline
$V_{NN}$ &$k_F$ [{\rm fm}$^{-1}$] & $V_0$[fm] & $V_2$[{\rm fm}] 
& $V_4$[{\rm fm}] & Sum & $\xi$\\
\hline
CDBonn   & 1.2  & -2.053  & 3.169 & -1.801 & -0.445  & 0.434  \\
HCSW01   &     & -2.001  & 2.865 & -1.402  & -0.441  & 0.439  \\ 
HCSW02   &     & -1.904  & 2.373 & -0.999 & -0.440  & 0.440  \\ 
HCSW03   &     & -1.893  & 2.261 & -0.825 & -0.440  & 0.439  \\ 
HCSW01   & 1.0  & -2.102 &  2.202  & -1.070 & -0.526  & 0.442  \\ 
HCSW01   & 1.4  & -1.945  & 3.584 & -1.983  & -0.375  & 0.443  \\ 
\hline
\end{tabular}
\label{table:coeffexp}
\end{table}

\begin{figure}[h]
\scalebox{0.42}{
\includegraphics[angle=-90]{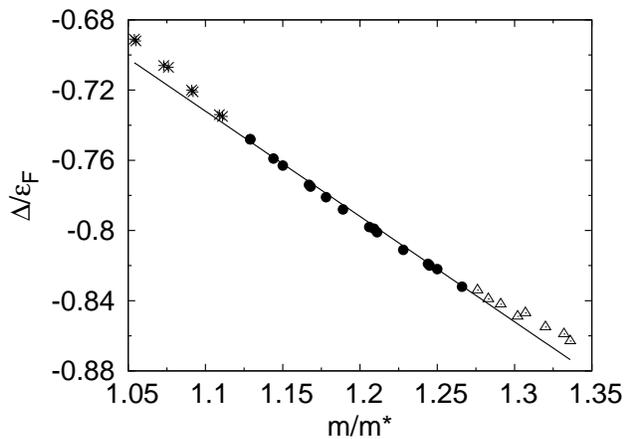} 
}
\caption{\label{linear} MSHF effective mass $m^*$ and well depth $\Delta$
at the unitary limit. Results calculated from the unitarity potentials
are denoted by `star', `dot' and `triangle' respectively for $k_F$
in the ranges (0.8-0.9), (0.9-1.4) and (1.4-1.5) {\rm fm}$^{-1}$. 
The solid line represents the linear expression 
 Eq.(18) with $\xi$=0.44. See text for other explanations.}
\end{figure}

 By far, the best numeric estimates of the ratio $\xi$ have been
obtained from the variational quantum-Monte-Carlo (QMC) calculations
(see e.g. \cite{giorgini} and references quoted therein). It would be
useful and of interest to compare our ring-diagram
and MSHF calculations with the QMC ones. 
 As reviewed in \cite{giorgini}, superfluid-QMC calculations are
generally  based on the  variation principle
$\delta [\langle \Phi_{BCS}|\Psi_J^{\dagger}H\Psi_J|\Phi_{BCS} \rangle/  
\langle \Phi_{BCS}|\Psi_J^{\dagger}\Psi_J|\Phi_{BCS} \rangle] =0,$
where  $\Phi_{BCS}$ is a BCS-paired trial wave function of definite 
number of particles and $\Psi _J$ is a r-space Jastrow correlator
whose inclusion  depends on the interaction
employed; its inclusion is necessary when an interaction with a `hard-core'
repulsion is employed \cite{gezerlis} but not so for smooth potentials 
without `hard core' \cite{carlson,chang04,giorgini}.
Clearly pairing is  incorporated in superfluid-QMC
by using a BCS-paired trial wave function. As discussed in section II, 
we emphasize pairing differently, namely by including only 
the boson-forming ring diagrams to all orders (both forward and 
backward ladder diagrams for a pair of fermions interacting 
any number of times), and in this way our all-order ring-diagram calculation
 treats the system as a collection of quasi bosons.
Despite this difference, it is encouraging that our result of 
$\xi \simeq$0.44 is satisfactorily close to the superfluid-QMC value
of $\sim 0.40$ to $\sim 0.44$ \cite{giorgini,gezerlis}.   
It may be pointed out that when we include only the first order 
ring diagram, our ring-diagram calculation reduces to the usual HF
calculation. For $\Lambda=2.1 fm^{-1}$ and $k_F=0.8 fm^{-1}$ 
we have carried out  HF 
calculations (i.e. including only
the 1st-order ring diagram) and obtained $\xi_{HF}\simeq$0.55, for all the
four unitarity potentilas mentioned earlier. This $\xi _{HF}$ is 
significantly larger than the corresponding all-order-ring result of 0.44. 
Normal-QMC calculations (where the trial wave function is a 
closed Fermi sea) have given $\xi \simeq$0.54 \cite{carlson03,giorgini}.
We believe that our 1st-order ring calculation is
similar to that of the normal-QMC 
while our all-order ring-diagram one is similar to
that of the superfluid-QMC (where the trial wave function is $\Phi _{BCS}$). 
Further studies  of the above possible connections are needed.

We have discussed in section II that at the unitary limit the MSHF
s.p. potential has a special property, namely its effective mass $m^*$
and well depth $\Delta$ satisfy a linear constraint dependent on the
 unitary ratio $\xi$ (see Eq.(18)).
Here let us check how well is this constraint obeyed by our results.
Similar to Fig.5, we have calculated $m/m^*$ and $\Delta/\varepsilon _F$
using the CDBonn and HCSW unitarity potentials using a wide range of
 $k_F$ from 0.8 to 1.5 {\rm fm}$^{-1}$. Our results are plotted in Fig.7,
and as seen most of our data are located near the straight line
corresponding to $\xi$=0.44. In the figure, the results for $k_F$
in the range of (0.9-1.4){\rm fm}$^{-1}$ are quite close to the $\xi$=0.44
straight line. For $k_F$ outside this range, they are slightly off
the line, which is consistent with the results of Fig. 5.

 It is promising that at the unitary limit the  above MSHF 
framework has given 
rather satisfactory results for the ratio $\xi$ ($\sim 0.44$), 
in close agreement with  both the superfluid-QMC results 
\cite{gezerlis,giorgini}
and those from our all-order ring diagram calculations. 
In addition, this method is fairly simple
and transparent; as indicated by Eq.(15) $\xi$ is  given by a 
simple integral and the HF potentials at the unitary limit
satisfy the constraint of Eq.(19). It may be mentioned, however, 
that this MSHF is only a limited effective theory; it is the
$\Lambda=k_F$ limit of the ring-diagram formalism. In this limit
the model space is one dimensional, and the model-space ground state
wave function is $| k_F\rangle$, the closed Fermi sea.
The BCS pairing gap 
$\Delta_{BCS}$ is an important quantity, but this information is not
contained in the MSHF effective theory.
$\Delta_{BCS}$ is the lower bound for
 the lowest excitation energy of the many-body system.   
 MSHF, however, is a one-dimensional effective theory  which can describe
only the ground state of the system; it provides no information about 
excited states. The effective MSHF theory is not
capable to provide information about $\Delta_{BCS}$,
even though the MSHF wave function $|k_F \rangle$
is completely $(\vec k \uparrow, -\vec k \downarrow)$ paired.  
To calculate $\Delta_{BCS}$, one may need to employ the number-nonconserving
Green's function framework \cite{fetter,suyangkuo} which has been commonly used
in supercoductivity (superfluidity) calculations. (The Green's function
method used in the present work is based on a number-conserving formalism, 
and is not suitable for calculating $\Delta _{BCS}$.) Using several Skyrme
 effective interactions, Su. et al. \cite{suyangkuo} have performed such
number-nonconserving BCS calculations for nuclear matter. We plan to
repeat their BCS calculations using instead the RG $V_{low-k}$ interactions,
and  will report our results in a separate publication.

\section{SUMMARY AND CONCLUSION} 
Using several different unitarity potentials, defined as having
infinite (very large) scattering length $a_s$, we have calculated the
ground-state energy ratio $\xi\equiv E_0/E_0^{free}$ for neutron
matter over a wide range of densities. A main purpose of our study
was to check if the ratios so obtained are `universal' 
in the sense that they have a common value (or nearly so),
 independent of the details of the potentials as long as
they have $a_s \rightarrow \pm \infty$. We have used four unitarity
potentials: One is the unitarity CDBonn potential whose meson exchange
parameters are slightly tuned so that its $a_s$ becomes very large 
(-12070 {\rm fm}). The other three are square-well `box' potentials with
both hard-core repulsion and exterior attraction. The $a_s$ of these
box potentials can be obtained analytically, making it easy to construct
many unitarity box potentials of different ranges and depths. 

We have calculated $\xi$ using two methods: a ring-diagram method and a 
model-space Hartree-Fock (MSHF) method.  An important step in both methods
is the transformation of the unitarity
potentials, which have strong short-range repulsions, 
into  low-momentum
interactions $V_{low-k}$ which are smooth potentials, convenient
for many-body calculations. The transformation is carried out using
a renormalization group method which preserves low-energy phase shifts
and the scattering length. We have found that the ratios $\xi$ given
by the ring-diagram and MSHF methods are practically identical
over a wide range of densities  (from $\sim 0.02$ to $\sim 0.09$
{\rm fm}$^{-3}$). This is an indication that the particle-particle hole-hole
correlations (as shown by diagram (c) of Fig.1) are not important
for neutron matter. Such correlations are included  in the ring-diagram
method but not in MSHF. The effect of the particle-particle
ladder correlations (as shown by diagram (b) of Fig.1) are included
in both. The MSHF method is considerably simpler than 
the ring-diagram one. It may provide a promising method for neutron
matter. Further study of this method for neutron matter may be
useful and of interest.

The CDBonn and the three box unitarity potentials are very different
 from each other, but the ratios $\xi$ predicted by them using either of the
above methods are all remarkably
close to 0.44, over a wide range of densities. This is 
indeed a rather interesting and surprising  result, indicating that 
$\xi$ may be perceived a universal constant. For the MSHF case, this result is related to
some special properties of the $V_{low-k}$ interaction at the unitary limit.
 We have found that the $V_{low-k}$
interaction for MSHF can be accurately simulated by a
low-order momentum expansion of the form [$V_0+V_2(k/k_F)^2+V_4(k/k_F)^4$],
and at the unitary limit the strength sum $(V_0/3+V_2/10+3V_4/70)$
calculated from the above potentials satisfies  the
 constraint of Eq.(20) very well, for a wide range of $k_F$. It is also
found that at the unitary limit the parameters $m^*$ and $\Delta$ of
the MSHF mean field potentials given by the 
above potentials all obey a linear constraint satisfactorily.
In conclusion, we believe that  our results provide strong numerical support
to the conjecture that the ratio $\xi$ is universal.

{\bf Acknowledgement}
We thank G.E. Brown and E. Shuryak for many helpful discussions.
 This work is supported in part by the U.S. 
Department of Energy under Grant Nos. DE-FG02-88ER40388 
and DE-FG02-03ER41270 (R.M.), and the U.S.
National Science Foundation under Grant No. PHY-0099444.

\end{document}